\documentclass[UTF8,letterpaper,english,aps, prb, reprint, superscriptaddress]{revtex4-1}
\usepackage[T1]{fontenc}
\usepackage[latin9]{inputenc}
\usepackage{amsmath}
\usepackage{amssymb}
\usepackage{bbold}
\usepackage{graphicx}
\usepackage[colorlinks,linkcolor=blue,anchorcolor=blue,citecolor=blue]{hyperref}
\usepackage{float}
\usepackage{caption}
\usepackage{placeins}

\makeatletter
\pdfpageheight\paperheight
\pdfpagewidth\paperwidth

\PassOptionsToPackage{position=top}{subfig}

\hypersetup{
	colorlinks = true,
	allcolors = blue
}

\usepackage{times}
\makeatother
\usepackage{babel}
\begin{document}
\title{Classifying extended, localized and critical states in quasiperiodic lattices via unsupervised learning}
\author{Bohan Zheng}
\thanks{These authors contributed equally to this work.}%
\affiliation{School of Computer Science and Technology, School of Software, Nanjing University of Posts and Telecommunications, Nanjing, 210023, China}%

\author{Siyu Zhu}
\thanks{These authors contributed equally to this work.}%
\affiliation{Institute of Quantum Information and Technology, Nanjing University of Posts and Telecommunications, Nanjing 210003, China}%

\author{Xingping Zhou}
\thanks{zxp@njupt.edu.cn}
\affiliation{Institute of Quantum Information and Technology, Nanjing University of Posts and Telecommunications, Nanjing 210003, China}%

\author{Tong Liu}
\thanks{t6tong@njupt.edu.cn}
\affiliation{Department of Applied Physics, School of Science, Nanjing University of Posts and Telecommunications, Nanjing 210003, China}%


\begin{abstract}
Classification of quantum phases is one of the most important areas of research in condensed matter physics. In this work, we obtain the phase diagram of one-dimensional quasiperiodic models via unsupervised learning. Firstly, we choose two advanced unsupervised learning algorithms, Density-Based Spatial Clustering of Applications with Noise (DBSCAN) and Ordering Points To Identify the Clustering Structure (OPTICS), to explore the distinct phases of Aubry-Andr\'{e}-Harper model and quasiperiodic p-wave model. The unsupervised learning results match well with traditional numerical diagonalization. Finally, we compare the similarity of different algorithms and find that the highest similarity between the results of unsupervised learning algorithms and those of traditional algorithms has exceeded 98\%. Our work sheds light on applications of unsupervised learning for phase classification.

\vspace{0.5cm}
\noindent $ \textbf{Keywords:} $ quantum phase, quasiperiodic, machine learning

\vspace{0.5cm}
\noindent $ \textbf{PACS:} $ 71.23.Ft, 71.10.Fd, 71.23.An
\end{abstract}

\maketitle

\section{INTRODUCTION}\label{n1}
Anderson localization \cite{r1}, the suppression of wave diffusion in disordered media, is ubiquitous across many areas of classical and quantum physics. Numerous experimental demonstrations have been reported in various systems, such as photonic systems \cite{r2,r3} and ultracold atoms \cite{r4}. Currently, the theoretical framework of Anderson localization is well established. The scaling theory predicts no delocalization in one- and two-dimensional systems, while three-dimensional (3D) systems can exhibit a localization-delocalization transition \cite{r5}. A threshold energy that separates extended from localized eigenstates is referred to as the mobility edge (ME) \cite{r6}. 

Quasicrystals also display novel localized phenomena \cite{r7,r8}. Unlike random systems, quasiperiodic systems exhibit localization transitions even in one dimension. A paradigmatic model is provided by the well-known Aubry-Andr\'{e}-Harper (AAH) model \cite{r9,r10}, where the localization-delocalization transition can be derived from a simple self-duality argument. Recently, significant interest has focused on finding low-dimensional quasicrystals that, analogous to the 3D Anderson model, display MEs separating extended and localized states \cite{r11,r12,r13,r14,r61,r62,r63,r64,r65,r66,r67,r68,r69,r70,r71,r72,r73}.

Traditional methods for distinguishing extended, localized, and critical states across these various models involve calculating a typical physical quantity, the inverse participation ratio (IPR) \cite{r15,r16,r17}. For a normalized wave function, the IPR of an extended state scales linearly as \(\frac{1}{L}\), vanishing in the thermodynamic limit \(L \to \infty\); the critical state approaches zero at a slower (power-law) rate, while the IPR of a localized state remains finite. Another method involves calculating the Lyapunov exponent \(\gamma\), which is defined as the divergence rate between neighboring lattice points. For an extended state, the amplitude between neighboring lattice points remains equal in the thermodynamic limit, resulting in \(\gamma = 0\); conversely, for a localized state, the amplitude decays exponentially, leading to \(\gamma > 0\). However, the single Lyapunov exponent cannot distinguish the extended state from the critical state, hence the numerical results of critical states are often ambiguous, and a exact definition of critical states is lacking. 

Remarkablely, a recent work \cite{r18} proposes an explicit criterion for precisely characterizing critical states, asserting that the Lyapunov exponents of critical states should simultaneously be zero in both position space and momentum space. In more physically-based language, critical states should exhibit anomalous delocalization transport and multifractal structure (non-extended and non-localized) in both position space and momentum space.
\begin{figure*}
  \centering
  \includegraphics[scale=0.96,width=0.96\textwidth]{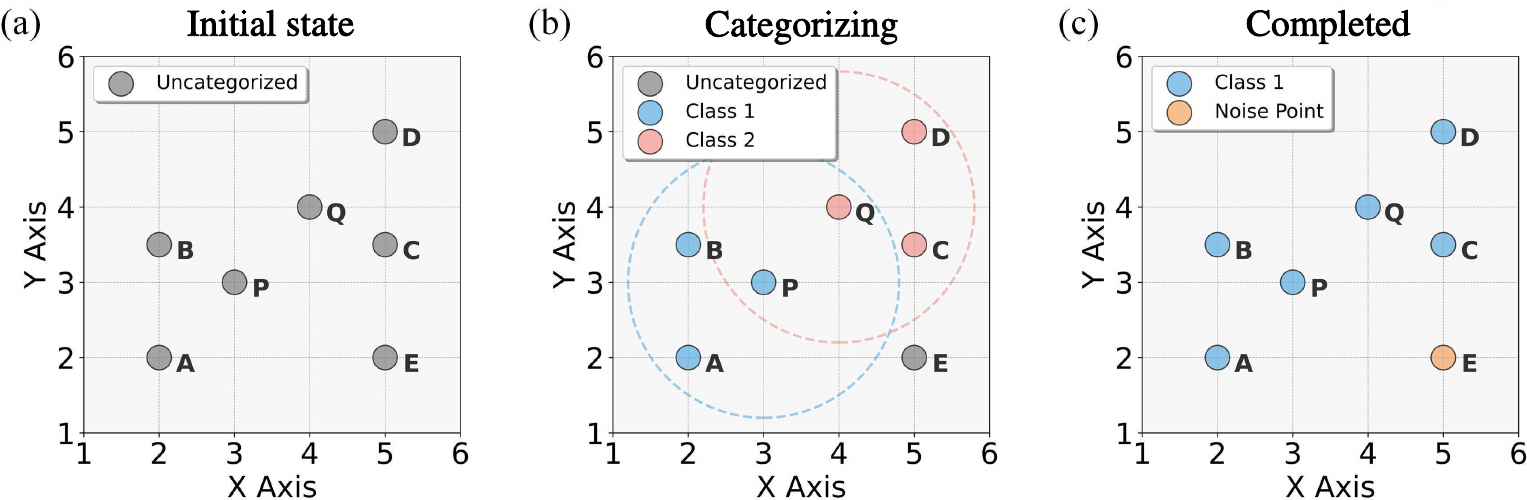}\\
  \captionsetup{width=0.96\textwidth, justification=raggedright, singlelinecheck=false}
  \caption{An example that demonstrates the process of the DBSCAN algorithm.}
  \label{fig1}
\end{figure*}

An interesting question arises: Are there alternative methods to distinguish different quantum states in disordered systems? The rise of machine learning \cite{r19,r20} research seems to answer this question, which has numerous applications in physical contexts, such as photonic structure design \cite{r24}, quantum many-body physics \cite{r25,r26,r27}, quantum computing, and chemical and material physics \cite{r28}, as well as topological phase classification \cite{r29,r30,r31,r32}. Unsupervised learning \cite{r33}, a significant branch of machine learning, can facilitate the data-driven construction of quantum states without prior knowledge.

In previous works, supervised learning methods \cite{r36,r37} have been employed to detect classical and quantum phase transitions \cite{r38,r40}, predict the phase diagrams of the long-range Harper model and the AAH model \cite{r41}, and address the edge learning problem of single event migration \cite{r42,r43}. However, many questions remain unresolved. For example, prior works have yet to derive eigenstates as markers through supervised learning. Notably, unsupervised learning algorithms do not require a training set; rather, the algorithms independently search for features within the data. This implies that phase diagrams can be obtained without traditional methods. Furthermore, previous algorithms \cite{r36,r37} have not effectively distinguished critical states from extended and localized states. In this work, we demonstrate the capacity of the unsupervised classification of three typical eigenstates in one-dimensional quasiperiodic models.

\section{DBSCAN AND OPTICS ALGORITHM}\label{n2}


First, let's introduce the principle of unsupervised learning algorithm. In practical calculations, we employ two widely-used algorithms, Density-Based Spatial Clustering of Applications with Noise (DBSCAN) \cite{r80,r33} and Ordering Points To Identify the Clustering Structure (OPTICS) \cite{r81,r33}, 
due to their robustness in handling complex data. DBSCAN, introduced by Martin Ester et al. in 1996, is capable of identifying clusters of arbitrary shapes without requiring the pre-specified clusters, making it particularly suitable for distinguishing localized states (high density) and extended states (low density). OPTICS, developed by Ankerst et al. in 1999, orders points based on their density reachability, allowing for a more detailed analysis of clustering structures, which is useful for detecting transitions between extended, localized, and critical states.



The inputs for both algorithms are n-dimensional vectors, which are treated as coordinate points in an n-dimensional space. The output consists of labels of these points. The algorithms rely on the distance between two points, commonly defined by the Euclidean distance, which is given by the formula: 
\begin{equation}\label{eq1} 
d = \sqrt {\sum\limits_{i = 1}^n {{{\left( {{x_{1,i}} - {x_{2,i}}} \right)}^2}} }, 
\end{equation} 
where ${x_{1,i}}$ and ${x_{2,i}}$ represent the coordinates of two points in the $i$-th dimension, respectively.

\begin{figure*}
  \centering
  \includegraphics[scale=0.96,width=0.96\textwidth,keepaspectratio=true]{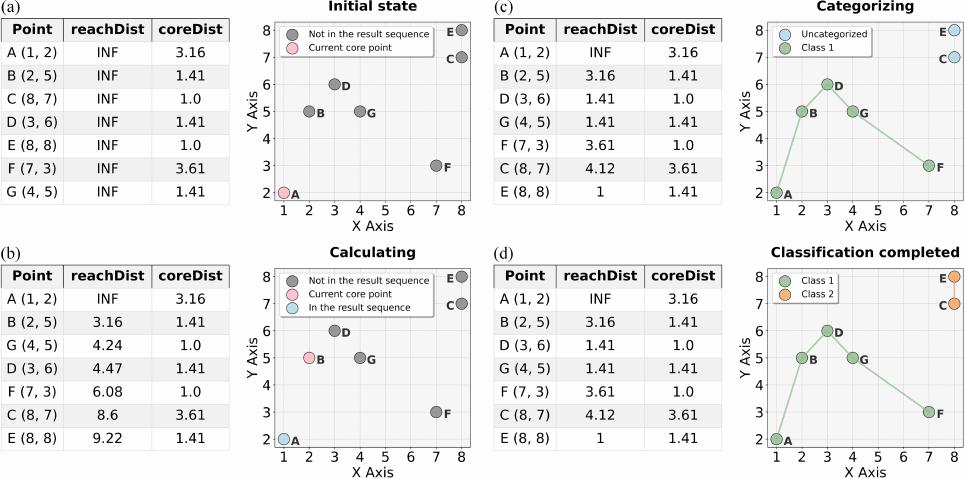}\\
  \captionsetup{width=0.96\textwidth, justification=raggedright, singlelinecheck=false}
  \caption{An example that illustrates the OPTICS algorithm process.}
  \label{fig2}
\end{figure*}

\subsection{DBSCAN algorithm}\label{n2_1}

The DBSCAN \cite{r80,r33} is effective for clustering data of arbitrary shapes, as its process is illustrated in Fig.~\ref{fig1}. DBSCAN categorizes data points into distinct clusters based on the distances between them, which is calculated using the Euclidean distance Eq.~(\ref{eq1}).

The core idea of the DBSCAN algorithm is to group together points that are density-connected---that is, connected with sufficient density---while marking those that are not part of any cluster as noise. The concept of "density-connected" is defined by two parameters: $EPS$ (the neighborhood radius) and $MinPoints$ (the minimum number of points required to form a cluster). A point is classified as a core point if it has at least $MinPoints$ neighbors within its $EPS$ radius. If a point $q$ lies within the $EPS$ radius of a core point $p$, then $q$ is considered directly density-reachable. Subsequently a point $r$ is said to be density-connected with $p$ if there exists a chain of directly density-reachable points connecting them. 

An intuitive example is illustrated in Fig.~\ref{fig1}, the parameters are set to $MinPoints = 3$ and $EPS = 1.8$. In Fig.~\ref{fig1}(b), point $P$ has points $Q$, $A$, and $B$ within its $EPS$ radius, while point $Q$ has points $P$, $C$, and $D$ in its neighborhood, making both $P$ and $Q$ core points. These core points and their neighbors form a cluster. Since $P$ and $Q$ are density-connected, their clusters merge in Fig.~\ref{fig1}(c). Point $E$ does not belong to any cluster and is classified as noise.

\subsection{OPTICS algorithm}\label{n2_2}

In contrast to DBSCAN, which only classifies points that exceed a fixed density threshold into clusters, OPTICS \cite{r81,r33} is capable of identifying clusters with varying density levels, grouping points with similar densities into the same cluster. Fig.~\ref{fig2} illustrates its process.

The core idea of the OPTICS is to generate a sequence that reflects the positional relationship between data points, and cluster the data points based on this sequence. Each data point in the sequence has two attributes: the core distance ($coreDist$), defined as the distance to its $MinPoints$-th nearest neighbor, representing the density around the point, and the reachability distance ($reachDist$), which reflects the distance between the point and preceding points in the sequence. 

The specific sequence generation process is as follows. Initially, the algorithm computes the $coreDist$ for each point and sets all $reachDist$ values to infinity. A point is then randomly selected as the first core point [Fig.~\ref{fig2}(a)]. In each iteration, the new $reachDist$ value is calculated as the larger between the distance from point $i$ to the current core point $p$ and the $coreDist$ of $p$. If this new value is smaller than the previous $reachDist$, it is updated accordingly. The point with the smallest $reachDist$ is chosen as the next core point, while the previous core point is added to the result sequence and excluded from further calculations [Fig.~\ref{fig2}(b)].

After the iterations, clusters are determined based on the result sequence and the threshold $EPS$ [Fig.~\ref{fig2}(c) and (d)]. If a point's $reachDist$ is smaller than $EPS$, it is grouped with the previous point in the same cluster. If its $reachDist$ is greater than $EPS$ but its $coreDist$ is smaller than $EPS$, it starts a new cluster. Otherwise, it is classified as noise.

For instance, Fig.~\ref{fig2} illustrates seven points with $MinPoints = 1$ and $EPS = 4$. Initially, point $A$ is chosen as the core point [Fig.~\ref{fig2}(a)], and after the first round of $reachDist$ calculations, point $B$ is selected as the next core point, with $A$ being added to the result sequence [Fig.~\ref{fig2}(b)]. The process continues until all points are classified. Point $A$ forms a new cluster, and points $B$, $D$, $G$, and $F$ are included in $A$'s cluster, while points $C$ and $E$ form a separate cluster [Fig.~\ref{fig2}(c) and (d)].

\section{TWO TYPICAL MODELS}\label{n3}
In various quasiperiodic systems, the AAH model and the quasiperiodic p-wave model are two representative models due to their display of rich quantum states: extended, localized, and critical states.

\subsection{Aubry-Andr\'{e}-Harper model}\label{n3_1}
Quasiperiodic systems\cite{r44,r45,r46,r47,r48,r51} exhibit quasiperiodic structures rather than random distributions, such as the Fibonacci lattice model, the Thue-Morse lattice model, and the well-known AAH model. The lattice Hamiltonian for the AAH model is given by:
\begin{equation}\label{eq2}
\hat H =  - \sum\limits_{i = 1}^{L - 1} {t(\hat c_i^\dag {{\hat c}_{i + 1}} + H.c.)}  + \sum\limits_{i = 1}^L {V\cos (2\pi \alpha i + \phi ){{\hat n}_i}}.
\end{equation}
Here, $\hat c_i^\dag$ and $\hat c_{i + 1}$ represent the fermion creation and annihilation operators, respectively; $\hat n_i = \hat c_i^\dag \hat c_i$ is the particle number operator; and $L$ denotes the total number of lattice points. The term $V\cos (2\pi \alpha i + \phi )\hat n_i$ represents the quasiperiodic potential field. The parameter $\alpha = \frac{{\sqrt{5} - 1}}{2}$ is an irrational number. Without loss of generality, we choose the phase factor $\phi = 0$ and set $t = 1$ as the energy unit for numerical calculations.

\begin{figure*}
  \centering
  \includegraphics[scale=0.96,width=0.96\textwidth,keepaspectratio=true]{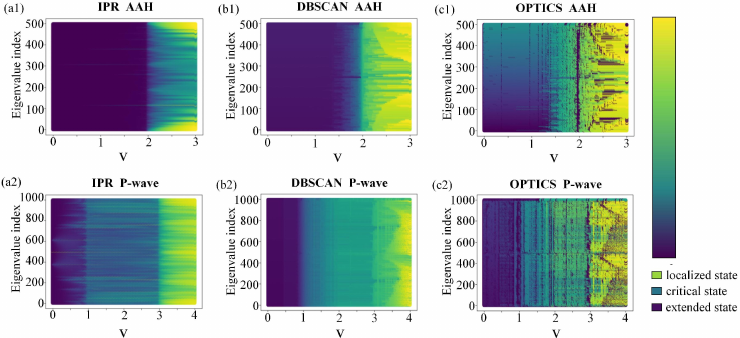}\\
  \captionsetup{width=0.96\textwidth, justification=raggedright, singlelinecheck=false}
  \caption{Figures of the IPR for the AAH model (a1) and the quasiperiodic p-wave model (a2). Clustering results of the eigenstates for (b1) the AAH model and (b2) the quasiperiodic p-wave model using the DBSCAN algorithm with parameters (b1) $EPS = 0.2$ and $Minpoints = 1$, and (b2) $EPS = 0.5$ and $Minpoints = 1$. Clustering results of the eigenstates for (c1) the AAH model and (c2) the quasiperiodic p-wave model using the OPTICS algorithm with parameters (c1) $EPS = 0.2$ and $Minpoints = 2$, and (c2) $EPS = 0.8$ and $Minpoints = 2$.}
  \label{fig3}
\end{figure*}

In quantum disordered systems, the inverse participation ratio (IPR) is a quantity typically used to characterize the localized and extended properties of eigenstates. The variation of IPR in the AAH model with respect to the disorder potential strength $V$ within the range (0, 3) is shown in Fig.~\ref{fig3}(a1). The brightness of the color represents the value of the IPR, indicating that brighter colors correspond to larger IPR values. This analysis clearly reveals a sudden change at $V = 2t$, where all eigenstates of the Hamiltonian become critical states. Moreover, all eigenstates of the Hamiltonian are extended when $V < 2t$, whereas they are localized when $V > 2t$.

\subsection{Quasiperiodic p-wave model}\label{n3_2}
The Hamiltonian for a one-dimensional p-wave superconducting pairing model in a quasiperiodic lattice is given by:
\begin{equation}\label{eq4}
H = \sum \left[ -t \hat c_i^\dag \hat c_{i + 1} + \Delta \hat c_i \hat c_{i + 1} + H.c. + V_i \hat n_i \right],
\end{equation}
where \( V_i = V \cos(2\pi \alpha i) \). When \( \Delta = 0 \), the model in Eq.~(\ref{eq4}) reduces to the AAH model in Eq.~(\ref{eq2}). When \( \Delta \neq 0 \) and \( V \) increases, this model exhibits a transition from a topological superconducting phase to an Anderson localized phase at \( V = 2|t + \Delta| \). Moreover, the model in Eq.~(\ref{eq4}) exhibits a large number of critical states.

The numerical phase diagram of IPR for this model \cite{r52,r60} is shown in Fig.~\ref{fig3}(a2). In the region where \( V < 2|t - \Delta| \), all eigenstates of the system are extended states; in the region \( 2|t - \Delta| < V < 2|t + \Delta| \), all eigenstates are critical states; and in the region \( V > 2|t + \Delta| \), all eigenstates are localized states.


\subsection{Simulation results}\label{n3_3}
By applying the DBSCAN and OPTICS algorithms, We perform numerical simulations on these two models. For the AAH model (Eq.~\ref{eq2}), we set the total number of lattice points $L = 500$ to obtain the eigenvector of the Hamiltonian. Similarly, for the quasiperiodic p-wave model (Eq.~\ref{eq4}), we set $L = 1000$. These eigenvectors are put into the two algorithms separately, where the parameters $EPS$ and $MinPoints$ are adjusted. The output is a list of clustering labels, indicating the category to which each eigenvector belongs.

The clustering results for the AAH and quasiperiodic p-wave models are shown in Fig.~\ref{fig3}(b1), (b2), (c1) and (c2), respectively. The horizontal axis of each figure represents the disorder potential strength $V$, while the vertical axis represents the index of eigenvalues for a given $V$. Each point corresponds to an eigenvector, and different colors represent the distinct categories produced by the clustering. 

As shown in Fig.~\ref{fig3}, the classification of eigenstates is clearly visible. In Fig.~\ref{fig3}(b1) and (c1), the region where $V < 2$ is darker, representing the extended state; $V > 2$ corresponds to the localized state, and $V = 2$ marks the critical transition point between two phases. In Fig.~\ref{fig3}(b2) and (c2), $V < 1$ indicates the extended state, $V > 3$ represents the localized state, and the region $1 < V < 3$ corresponds to the critical state. Notably, the OPTICS algorithm distinctly identifies the critical state of the AAH model at $V = 2$ in Fig.~\ref{fig3}(c1).

To explain more specifically how DBSCAN and OPTICS algorithms classify extended and localized states, we demonstrate their frameworks for quantum disordered systems. DBSCAN groups together eigenstates with high-density, while marking those in lower-density regions as outliers. Hence DBSCAN can identify extended states, which are more uniformly spread across the lattice, by forming large and continuous clusters. On the other hand, localized states, where the wave function is confined to a smaller region, result in more compact and isolated clusters. OPTICS, similar to DBSCAN, focuses on ordering eigenstates based on the density reach-ability. This approach allows OPTICS to detect more subtle transitions in quantum systems, such as critical states, as the density structure captured by OPTICS reflects the gradual transition from localized to extended states, and critical states lie at the boundary between these phases.

The primary advantage of machine learning methods like DBSCAN and OPTICS over traditional methods lies in the data-driven nature. Machine learning algorithms can automatically detect patterns and classify states without requiring prior knowledge of the system. This capability makes them particularly effective for identifying complex phase transitions, including critical states, which are challenging by the IPR simulation. Moreover, machine learning methods react well for larger data-sets, allowing for more efficient classification in systems with a large number of eigenstates.

\section{SIMILARITY}\label{n4}
In this section, we employ the difference hash algorithm to calculate the similarity between traditional methods [Fig.~\ref{fig3}(a)] and machine learning methods [Fig.~\ref{fig3}(b) and (c)], namely the similarity between the IPR results and the clustering results.

The difference hash algorithm treats a figure as a two-dimensional signal composed of various frequency components. High-frequency components correspond to regions with significant brightness variations between adjacent pixels, providing detailed information about the image. In contrast, low-frequency components represent areas with minor brightness variations, capturing the general structure of the image. The algorithm reduces the image size to filter out high-frequency components and computes the hash values by focusing on the low-frequency components. If a pixel is brighter than the following pixel, the corresponding hash bit is set to 1; otherwise, it is set to 0. The similarity between two images is then determined by comparing their hash values. The similarity $Q$ is given by the formula:
\begin{equation}\label{eq5}
Q =  1 - \frac{\sum\limits_i^N {(h_{1,i} + h_{2,i} + 1)\bmod 2}}{N},
\end{equation}
where $N$ is the number of pixels in the image, and $h_{1,i}$ and $h_{2,i}$ are the $i$-th bits of hash values, respectively. 

\begin{table}[htb]
\caption{Similarity results for different algorithms. \label{table1}}
\begin{ruledtabular}
\begin{tabular}{c c c}
Similarity & DBSCAN & OPTICS  \\ \hline
AAH & 98.40\% & 90.63\%   \\
quasiperiodic p-wave &95.30\%&62.50\% \\
\end{tabular}
\end{ruledtabular}
\end{table}

Similarity results for different algorithms are illustrated in Table~\ref{table1}, which indicate that the DBSCAN algorithm significantly surpasses the OPTICS algorithm. For both the AAH model and the quasiperiodic p-wave model, the DBSCAN algorithm achieves a similarity of over 95\% compared to traditional methods, with a peak value of 98.4\%. In contrast, the OPTICS algorithm only attains approximately 90\% and 62\% similarity for the two models, respectively. Additionally, the unsupervised learning results for the AAH model are notably better than those for the quasiperiodic p-wave model. This is reasonable, as the AAH model deals with a single-particle problem, while the quasiperiodic p-wave model is a mean-field approximation of a strongly correlated system. The latter's increased complexity and reduced robustness to machine learning algorithms explain the lower performance.

\section{CONCLUSION}\label{n5}
In this work, we investigate the capability of unsupervised learning algorithms to extract information about distinct phases in various quasiperiodic systems. Specifically, we employ two unsupervised learning algorithms, DBSCAN and OPTICS, to classify the extended, localized, and critical states in the AAH model and the quasiperiodic p-wave model. While previous studies have focused on supervised learning algorithms, we demonstrate that unsupervised learning algorithms can accurately reproduce phase diagrams in close agreement with traditional numerical methods. Furthermore, we apply the difference hash algorithm to quantify the similarity between the unsupervised learning phase diagrams and the traditional numerical phase diagrams. Our results show that the DBSCAN algorithm is particularly effective for exploring quasiperiodic systems. Additionally, DBSCAN is not only applied to single-particle problems but also effectively describes the mean-field approximation of strongly correlated systems. Thus, a potential extension of DBSCAN is to distinguish many-body wave functions across various phases in many-body interacting systems. This work provides a valuable demonstration of unsupervised learning in classifying different states of matter and highlights its potential for phase classification.


\begin{acknowledgments}
This work was supported by the Natural Science Foundation of Nanjing University of Posts and Telecommunications (Grant No. NY223109, NY220119, NY221055), China Postdoctoral Science Foundation (Grant No. 2022M721693), and National Natural Science Foundation of China under (Grant 12404365).
\end{acknowledgments}




\end{document}